\documentclass[aip,jcp,superscriptaddress,twocolumn]{revtex4}
\usepackage{natbib}
\usepackage{color}
\usepackage{graphicx}       
\usepackage{dcolumn}        
\usepackage{amsmath}
\usepackage{amssymb}
\usepackage{epsf}
\usepackage{graphics}
\usepackage{rotating}
\usepackage{bm}
\usepackage{enumerate}  

\begin{document}

\title{Using Bayes formula to estimate rates of rare events in transition path sampling simulations}

\author{Pierre Terrier}
\affiliation{CEA, DEN, Service de Recherches de M\'etallurgie Physique, UPSay, F-91191 Gif-sur-Yvette, France}

\author{Mihai-Cosmin Marinica}
\affiliation{CEA, DEN, Service de Recherches de M\'etallurgie Physique, UPSay, F-91191 Gif-sur-Yvette, France}

\author{Manuel Ath\`enes}
\affiliation{CEA, DEN, Service de Recherches de M\'etallurgie Physique, UPSay, F-91191 Gif-sur-Yvette, France}

\begin{abstract}
Transition path sampling is a method for estimating the rates of rare events in molecular systems based on the gradual transformation of a path distribution containing a small fraction of reactive trajectories into a biased distribution in which these rare trajectories have become frequent. Then, a multistate reweighting scheme is implemented to postprocess data collected from the staged simulations. Herein, we show how Bayes formula allows to directly construct a biased sample containing an enhanced fraction of reactive trajectories and to concomitantly estimate the transition rate from this sample. The approach can remediate the convergence issues encountered in free energy perturbation or umbrella sampling simulations when the transformed distribution insufficiently overlaps with the reference distribution. 
\end{abstract}

\maketitle

\section{Introduction}
 
The frequencies or rates of thermally activated events are crucial parameters that control atomic transport in condensed matter at equilibrium and the long-term evolution of many systems driven out of equilibrium. Not surprisingly, considerable effort has been devoted in the last decades to designing  efficient molecular simulation methods for computing these rates. 
When the typical durations of the activated events are much shorter than the mean inter-event times, each event can be schematized as a transition of the system from a reactant basin ($a$) to a product basin ($b$) and the entire transiton paths can be studied using molecular dynamics (MD) simulations. So as to monitor $a$-to-$b$ transitions in MD, the practitioner is first faced to the problem of defining the suitable functions taking input values in the configuration space $\Omega$ and indicating whether a particular configuration belongs to one of the two basins of interest. For this purpose, we usually consider the indicator function 
$h_{a(b)}: q\in \Omega \rightarrow \left\{ 0 , 1\right\}$ whose output value is 1 
if the input is in subset $a(b)$ and 0 elsewhere. This enables us to formalize an $a$-to-$b$ time-correlation function as follows
\begin{eqnarray}
 C(t) & = & \frac{\int_0 ^{+\infty} h_a \left[ q (s) \right] h_b \left[ q(s+t) \right] ds }{\int_0 ^{+\infty} h_a \left[ q(s) \right] ds} 
\end{eqnarray}
in which $q(s)$ is the system configuration at time $s$. 
When the fast molecular relaxation and the overall residence time within basin $a$ occur on well-separated time scales, the derivative of the time-correlation function, $dC(t)/dt$, displays a transient plateau corresponding to the  phenomenological rate for transitioning from $a$ to $b$~\cite{chandler1978statistical,chandler:1987}. 
The accurate estimation of $C(t)$ is a challenging task because transition events are rare on the simulation timescale. This problem can be alleviated by the use of the transition path sampling method~\cite{dellago1999calculation,dellago:2002,elmatad:2010,geiger:2010,athenes:2012} (TPS) in which a bias is introduced to enforce~\cite{dellago1999calculation,dellago:2002} or favour~\cite{elmatad:2010,geiger:2010,athenes:2012} the sampling of the short trajectories transitioning from the product to the reactant basins. TPS writes the $a$-to-$b$ time-correlation function as a conditional expectation of the indicator function of basin $b$ at time $t$ over an ensemble of short trajectories starting at equilibrium in basin $a$ at time $t=0$,  
\begin{eqnarray}
 C(t) & = & \mathbb{E} \left\{ h_b \left[q(t)\right] | h_a\left[ q(0) \right] = 1 \right\}. \label{eq:correlation}
\end{eqnarray}
Let now write $\mathcal{O}$ the functional $h_b[q(t)]$ associated with paths starting in basin $a$ at time $t=0$. In principle, the time-correlation function in~\eqref{eq:correlation} can be estimated from the information contained in a set of $M$ trajectories of duration $t$, after correcting for the simulation biases $\left\{ e^{-B^m} \right\}_{1 \leq m \leq M}$ using the standard reweighting scheme used in the umbrella sampling method~\cite{frenkel:2002,lelievre:2010}
\begin{equation} \label{eq:conventional}
 \widehat{\mathcal{O}}^{M} = \frac{\tfrac{1}{M}\sum_{m=1}^M \mathcal{O}^m e^{-B^m }} {\tfrac{1}{M}\sum_{m=1}^M  e^{- B^m}}. 
\end{equation}
The observation $\mathcal{O}^m$ is the value taken by the path observable $\mathcal{O}$ for the the $m$th trajectory. 
The standard reweighting scheme also relates to the free energy perturbation (FEP) method.~\cite{frenkel:2002,lelievre:2010} 
The logarithm of the denominator in Eq.~\eqref{eq:conventional} corresponds to the free energy difference between the reference ensemble of interest and the biased (perturbed) ensemble that is sampled. To obtain an accurate estimate using~\eqref{eq:conventional}, the biased sample must however contain typical data of the unbiased distribution in a significant proportion. Stated differently, the perturbed and unpertubed distributions should substantially overlap. 
If this condition is not met, then the associated free-energy difference is usually overestimated.~\cite{bennett:1976,frenkel:2002,lelievre:2010}

In practice, the biased distributions that contains the reactive trajectories differs substantially from the reference distribution that contains non-reactive trajectories almost exclusively. As a result, the state-to-state correlation functions that have been estimated in TPS simulations so far were obtained through staged transformations.~\cite{dellago:2002,athenes:2012} The biasing potential is gradually switched on and a simulation is performed at each stage of the switching protocol. Then, a postprocessing procedure (the weighted histogram analysis method~\cite{ferrenberg:1989} or the multistate Bennett acceptance ratio method~\cite{shirts:2008}) is used to combine the data from the multiple simulations and, based on accurate estimates of the successive free-energy differences, to eventually extract a reliable estimate of the desired observable~\cite{frenkel:2002,lelievre:2010}. 

In this article, we show how Bayes formula can be used to (i) estimate any path observable expectation, (ii) adaptively construct a biasing potential whose associated distribution contains trajectories that are both reactive and non reactive (iii) estimate the state-to-state correlation function from a single sample owing to good overlapping properties of the biased distribution. The article is organized as follows. The Bayesian expectation in the rare event context is derived in Sec.~\ref{theory}. 
We then compare the approach to standard reweighting in Sec.~\ref{sec:testbed}. The migration of a vacancy in a crystal serves as an illustration of the approach in Sec.~\ref{sec:diffusion}. 

\section{Estimating path observables \label{theory}}
\subsection{Path ensembles and probability distributions}

A path $z$ consists of a sequence of $L+1$ states: $z= \left\{ q_\ell,p_\ell \right\}_{0 \leq \ell \leq L}$ where 
$q_\ell$ and $p_\ell$ are the $\ell$th positions and $\ell$th momenta of state $x_\ell = \left\{ q_\ell, p_\ell \right\}$ at time $t=\ell \tau$ with $\tau$ being the timestep of the considered MD scheme. 
From the conditional probability to generate $z$ using the MD scheme given initial state $x_0$, we define a path action  by 
\begin{eqnarray}
 \mathcal{H} (z) = H(x_0) - \beta^{-1} \ln \mathrm{P_{MD}} ( z |x_0).  
\end{eqnarray}
where $H$ is the system Hamiltonian and $\mathrm{P_{MD}} ( z |x_0)$ is the probability to generate path $z$ knowing that the system at $t=0$ is state $x_0$. 
In addition, a bias $\mathcal{K}(\theta,z)$, controlled by an external parameter $\theta$, acts upon path $z$ by modifying its occurrence probability. 
Confining the positions $q_0$ of the initial state of the path to basin $a$, the biased path distribution given $\theta$ writes 
\begin{equation} \label{eq:condproba}
 \pi(z|\theta) =  h_a(q_0) \exp \left[ \mathcal{A} (\theta) - \beta \mathcal{H} (z) - \mathcal{K}(\theta,z) \right]
\end{equation}
where the function $\mathcal{A} (\theta)$ acts as a normalizing constant in path space, denoted by $\mathcal{Z}$: 
\begin{equation} \label{eq:plugging}
 \exp \left[-\mathcal{A}(\theta) \right] = \int_\mathcal{Z} \exp \left[ - \beta \mathcal{H}(z) - \mathcal{K}(\theta,z) \right] \mathcal{D}z
\end{equation}

In a first TPS protocol,~\cite{dellago:2002} referred to as \emph{confining protocol}, the bias aims at confining the trajectory endpoints into windows distributed along a reaction coordinate (RC), a function $\xi: \mathcal{Q} \rightarrow \mathbb{R}$ that is able to describe the transition pathway from basin $a$ to basin $b$.  Resorting to a family of indicator functions $h^\xi_\theta$, the conditional probability writes $\pi(z|\theta) =  h_a(q_0) h_\theta^\xi (q_L) \exp \left[ \mathcal{A} (\theta) - \beta \mathcal{H} (z) \right]$, entailing
\begin{eqnarray} \label{p:confining}
 \mathcal{K} (\theta,z) & = &  \begin{cases} 0 &\mbox{if } h^\xi_\theta (q_L ) =1, \\
+\infty & \mbox{otherwise. } \end{cases} 
\end{eqnarray}
In Eq.~\eqref{p:confining}, paths ending outside window $h^\xi_\theta$ have zero probability~\eqref{eq:condproba} and $\theta$ usually takes values in a finite integer set $\Theta=\{ 0,1,2,\cdots, \theta_{\mathrm{max}} \}$. Window $h^\xi_0$ ideally contains the whole phase space : $\forall q \in \mathcal{Q}$, the configuration space, we have $h^\xi_0(q)=1$. Hence $\mathcal{K} (0,z)$ is always zero and $\pi(z| 0)$ corresponds to the unbiased path probability distribution. On the other end, window associated with $\theta_\mathrm{max}$ maps basin $b$, i.e.  $h^\xi_{\theta_{\mathrm{max}}}=h_b$. This implies that $\pi(z| \theta_\mathrm{max})$ is the probability distribution of the transition path ensemble and that the relation $C(L\tau) = \exp[-\mathcal{A}(\theta_\mathrm{max}) + \mathcal{A}(0)]$ holds. 

In a second TPS protocol, referred to as \emph{tilting protocol}, the bias acting upon the paths is the product of the external parameter and a path functional~\cite{athenes:2004,elmatad:2010,geiger:2010,athenes:2012}
\begin{equation} \label{p:tilting}
  \mathcal{K}(\theta,z) = \theta \mathcal{L}(z). 
\end{equation}
Since trajectory endpoints are not constrained, the fraction of reactive paths with respect to $\pi(z|\theta_\mathrm{max})$ distribution is no more equal to one. For  instance, in the recent set-up~\cite{athenes:2012} that will be considered below in Sec.~\ref{sec:diffusion}, $\mathcal{L}(z)$ is a non-positive functional that is all the more negative that the explored portions of the potential energy surface are more negatively curved. Because basins of attraction are separated from each others by mechanically unstable regions (corresponding to the negatively curved portions of the energy surface), the biasing approach is  able to increase the occurrence of reactive trajectories up to 10$\--40\%$, depending on the choice of $\theta_\mathrm{max}$.~\cite{athenes:2012} Although the equality between $C(L\tau)$ and $\exp[-\mathcal{A}(\theta_\mathrm{max}) + \mathcal{A}(0)]$ does  not hold anymore, the free energy difference $\mathcal{A}(\theta_\mathrm{max}) - \mathcal{A}(0)$ is again to be determined to estimate 
the unbiased correlation function.~\cite{athenes:2012} 

Unfortunately, the associated free-energy difference cannot be estimated directly through free energy perturbation from a single sample of unbiased trajectories because the correlation function is very small, nor from a sample of biased trajectories because typical non-reactive paths are not generated. That is why TPS constructs additional samples of trajectories, confined using intermediate windows~\cite{dellago:2002} or tilted using intermediate values for the external parameter.~\cite{athenes:2012} It finally resorts to a rematching procedure to extract the $\mathcal{A}(\theta_\mathrm{max}) - \mathcal{A}(0)$. This computational bottleneck results from the insufficient overlap between the biased and unbiased distributions, as illustrated below in Sec.~\ref{sec:testbed} within the tilting protocol~\eqref{p:tilting}. 

We now show that it is practically possible to construct a biased distribution exhibiting adequate overlapping properties. Assuming that $\Theta$ is $[0,\theta_\mathrm{max}]$ interval, 
this probability distribution reads 
\begin{eqnarray} \label{eq:pmarg}
 \bar{\mathrm{P}}_{\mathcal{A}} (z) =\frac{1}{\theta_\mathrm{max}}\int_\Theta \pi (z| d \theta) 
\end{eqnarray}
where the normalizing factor $1/\theta_\mathrm{max}$ corresponds to the uniform density of $\pi(z|\theta)$-distributions with respect to $\theta$. 
Evaluating $\bar{\mathrm{P}}_{\mathcal{A}} (z)$ still requires the accurate knowledge of the free energy $\mathcal{A}$, since this quantity enters the sampled probability density~\eqref{eq:pmarg} via~\eqref{eq:condproba}. However, the overall approach is greatly simplified owing to the use of the adaptive biasing force (ABF) technique~\cite{darve:2001,darve:2008,lelievre:2008,henin:2010} within the method of expanded ensembles.~\cite{lyubartsev:1992,iba:2001} The latter approach introduces an auxiliary biasing potential denoted by $A(\theta)$, assumed to be constant first, and considers the extended space $\Theta \cup \mathcal{Z}$ equipped with path action $A(\theta) - \mathcal{K}(\theta,z)$. The occurrence of $\theta$ in the expanded ensemble defines the marginal probability of $\theta$ 
\begin{equation} \label{marg_theta}
 \mathrm{P}_{A} (\theta ) \propto \exp \left[A(\theta) - \mathcal{A}(\theta)\right]. 
\end{equation}
This relationship is obtained after integrating over the path space $\mathcal{Z}$ and plugging~\eqref{eq:plugging}. It entails that, if $A$ is strictly equal to the free energy, then the marginal probability of $\theta$ is constant and equal to $1/\theta_{\mathrm{max}}$ in Eq.~\eqref{eq:pmarg}. This last feature suggests a way of constructing $\bar{\mathrm{P}}_{\mathcal{A}} (z)$ since the ABF method permits to adapt $A(\theta)$ on $\mathcal{A}(\theta)$ via the current estimate of its derivative $\mathcal{A}^\prime(\theta)$. Postponing the description of ABF technique to Sec.~\ref{subsec:adaptive}, we show how to estimate any observable expectation using Bayes formula.~\cite{cao:2014} 
This approach will require sampling the following marginal probability 
\begin{equation} \label{marg_z}
 \bar{\mathrm{P}}_{A} (z) = \int_{\Theta} \pi(z|\vartheta) \mathrm{P}_{A} (d \vartheta), 
\end{equation}
obtained after integrating $\vartheta \in \Theta$. This probability corresponds to the occurrence of any particular path in the expanded ensemble, irrespective of the value of $\theta$. 
Using the marginal probabilities~\eqref{marg_theta} and~\eqref{marg_z} and the conditional probability of $z$ given $\theta$, an analytical expression amenable to integration by numerical quadrature can be derived for the conditional probability of $\theta$ given $z$ from Bayes formula
\begin{eqnarray}
  \bar{\pi}_A (\theta | z ) & = & \frac{\pi(z| \theta) \mathrm{P}_A (\theta) }{  \bar{\mathrm{P}}_A (z)} = \frac{e^{A(\theta)-\mathcal{K}(\theta,z)}}{\int_\Theta e^{A(\vartheta)-\mathcal{K}(\vartheta,z)} d\vartheta} \label{theta|z}. 
\end{eqnarray}
Note that in~\eqref{theta|z} the conditional probability of $z$ knowing $\theta$ does not depend on the auxiliary biasing potential [see expression Eq.~\eqref{eq:condproba}], while the three other involved probabilities do. We  take advantage of this property and rearrange~\eqref{theta|z} so as to cast Bayes identity into a computationally useful form 
\begin{eqnarray} \label{eq:central}
 \pi (z | \theta ) & = & \frac{ \bar{\pi}_A (\theta | z )  \bar{\mathrm{P}}_A (z)}{\mathrm{P}_A (\theta)}. 
\end{eqnarray}
This relation will enable one to express the conditional expectation given $\theta$ as an expectation involving the conditional probabilities of $\theta$ given the set of paths sampled according to probability~\eqref{marg_z}. 

\subsection{Reweighting estimator based on Bayes formula~\label{subsec:BF_estimator}}

To show how this can be done, we first express the marginal probability of $\theta$ as an expectation involving the conditional probability of $\theta$ over the distribution of the marginal probability of $z$:  
\begin{eqnarray}
\pi (z | \theta )  & = & \frac{\bar{\pi}_A (\theta | z ) \bar{\mathrm{P}}_A (z)}{\int_{ \mathcal{Z}} \bar{\pi}_A (\theta | \tilde{z})  \bar{\mathrm{P}}_A (\mathcal{D}\tilde{z})} 
\end{eqnarray}
This equation can be used to cast $\mathbb{E}_\pi \left( \mathcal{O} | \theta \right) = \int_\mathcal{Z} \mathcal{O}(z)\pi(\mathcal{D}z|\theta)$, the conditional expectation of any path observable $\mathcal{O}(z)$ given $\theta$ into the following form  
\begin{eqnarray} \label{eq:ratio_a}
 \mathbb{E}_\pi \left( \mathcal{O} | \theta \right) & = & \frac{\int_{ \mathcal{Z}} \mathcal{O}(z) \bar{\pi}_A (\theta | z )  \bar{{\mathrm{P}}}_A (\mathcal{D} z) } {\int_{\mathcal{Z}} \bar{\pi}_A (\theta | z )  \bar{\mathrm{P}}_A (\mathcal{D} z)} 
\end{eqnarray}
Here, $\mathcal{O}(z)$ can be $h_b(q_\ell)$ the a-to-b time-correlation function or any path observable, such as $\mathcal{K}(\theta,z)$ or $\mathcal{L}(z)$ for instances. 

Let now assume that a sample of trajectories $\left\{ z^m \right\}_{1 \leq m \leq M}$ is constructed using a Monte Carlo scheme obeying detailed balance in the path ensemble of probability density $\bar{\mathrm{P}}_A$. Applying the ergodic theorem to the ensemble average ratio~\eqref{eq:ratio_a}, an estimator of the conditional expectation is 
\begin{equation} \label{eq:estimator}
 \hat{\mathbb{E}}^{\mathrm{B},M}_\pi \left( \mathcal{O} | \theta \right) = \frac{\tfrac{1}{M} \sum_{m=1}^M \mathcal{O}^m \bar{\pi}_A(\theta | z^m) } {\tfrac{1}{M} \sum_{m=1}^M \bar{\pi}_A(\theta | z^m)}
\end{equation}
where $\mathcal{O}^m=\mathcal{O}(z^m)$ and superscript $\mathrm{B}$ stands for Bayes formula. 
This reweighting approach has been previously used in a molecular context to extract free energy profiles.~\cite{cao:2014} It was termed adiabatic reweighting in reference to the dynamical decoupling that is involved in molecular dynamics simulations between the external parameter and the particle coordinates (See Ref.~\onlinecite{cao:2014}).

\subsection{Adaptive biasing force method~\label{subsec:adaptive}}

In ABF~\cite{darve:2001,darve:2008,lelievre:2008,henin:2010}, the derivative of the auxiliary biasing potential with respect to $\theta$ is adapted on the current estimate of $\mathcal{A}^\prime(\theta)$ the mean force along $\theta$. In the long term, $A$ converges to $\mathcal{A}$ (up to an additive constant) and $\mathrm{P}_A(\theta)$ becomes a uniform distribution. 
Differentiating $\mathcal{A}(\theta) $ is very simple mathematically within the tilting protocol~\eqref{p:tilting}. The mean force corresponds to the conditional expectation of $\mathcal{L}$ given $\theta$ 
\begin{equation} \label{eq:meanforce}
\mathcal{A}^\prime (\theta) = \int_\mathcal{Z} \partial_\theta \mathcal{K} (\theta,z) \pi(\mathcal{D}z|\theta) =\mathbb{E} (\mathcal{L} | \theta ). 
\end{equation}
In the application of Sec.~\ref{sec:diffusion}, $K$ replicas of the system will be simulated on a parallel computer architecture. We thus adapt the mean force $\mathcal{A}^\prime(\theta)$ for $\theta \in \Theta $ following Ref.~\onlinecite{cao:2014}
\begin{eqnarray}
 A^\prime_{\tilde{m}} (\theta) & = & \frac{\sum_{k=1}^K \sum_{m=1}^{\tilde{m}-1} \mathcal{L}(z^{km}) \pi_{A_{m}}(\theta | z^{km}) }
                {\sum_{k=1}^K \sum_{m=1}^{\tilde{m}-1} \pi_{A_{m}}(\theta | z^{km}) } \nonumber \\
                 & = & \frac{\sum_{k=1}^K \sum_{m=1}^{\tilde{m}-1} \mathcal{L}^{km}  \frac{ \exp \left[  A_m (\theta)- \theta \mathcal{L}^{km} \right] }{\int_\Theta \exp \left[  A_m (\vartheta)- \vartheta \mathcal{L}^{km} \right] d \vartheta} }
                {\sum_{k=1}^K \sum_{m=1}^{\tilde{m}-1}  \frac{  \exp \left[  A_m (\theta)- \theta \mathcal{L}^{km} \right] }{\int_\Theta \exp \left[  A_m (\vartheta)- \vartheta \mathcal{L}^{km} \right] d \vartheta} } \label{eq:abf_scheme}. 
\end{eqnarray}
The evaluation of the adaptive biasing force, a simple task compared to the evaluation of the interatomic forces, is shared by all cores. This is why the replica index $k$ is absent in $A^\prime_m$. The auxiliary potential is then integrated for the next step. 
After the preliminary ABF run, the auxilary biasing potential $A$ is frozen and the state-to-state correlation function is estimated in a subsequent run by setting $\mathcal{O}(z)$ to $h_b(q_\ell)$ ($0 \leq \ell \leq L$) and $\theta$ to 0. 
In this situation, the generic BF estimator writes  
\begin{equation} \label{eq:estimator_K}
 \hat{\mathbb{E}}^{\mathrm{B},KM}_\pi \left( \mathcal{O} | \theta \right) = \frac{\tfrac{1}{KM} \sum_{k=1}^{K} \sum_{m=1}^M  \mathcal{O}^{km} \bar{\pi}_A(\theta | z^{km}) } {\tfrac{1}{KM} \sum_{k=1}^{K} \sum_{m=1}^M \bar{\pi}_A(\theta | z^{km})}
\end{equation}
where $z^{km}$ refers to the $m$th path of the $k$th replica and $\mathcal{O}^{km}=\mathcal{O}(z^{km})$.

Note that ABF is primarily used to compute mean forces along one- and two-dimensional RC, which requires to evaluate the first and second derivatives of the RC with respect to $q$.~\cite{henin:2010,darve:2008} This task may be impractical in many circumstances, or even impossible when the RC is discrete for instance. Hence, it is not possible to implement ABF with the confining protocol~\eqref{p:confining} when the second derivative of $\xi$ is not defined. This limitation however does not restrict the scope of TPS because trajectory endpoints can be softly confined to basin $b$ by resorting to the tilting protocol~\eqref{p:tilting} and setting $\mathcal{L}(z)$ to $\xi(q_\ell)$, the RC value of the path end-point. This feature is illustrated in next subsection. 

\subsection{Confining path endpoints through the tilting protocol\label{subsec:confining}}

Let consider a generic model consisting of a particle evolving according to Brownian motion along a line (the momentum is omitted and the potential energy is constant). 
The position of the particle is obtained by integrating the corresponding overdamped Langevin equation: we have $ q_{\ell+1} = q_{\ell} + \sqrt{2\tau D} B_\ell $ where $\tau$ is the timestep, $D$ a diffusion coefficient and $B_\ell$ is a random variate drawn is the normal distribution of zero mean and unit variance. We set $a$ to $\{ 0 \}$ and and $b$ to $[1, + \infty ( $. 
This implies that  $q_0=0$ and $q_L$ is distributed according to the normal distribution $ q \rightarrow \sqrt{\omega / \pi }\exp(-\omega q^2) $ where $ \omega = (4 D L \tau)^{-1} $.  Integrating this distribution from 1 to $+\infty$ then yields the probability that a path ends at a position larger than $1$. This quantity defines the time-correlation $C(L\tau ) =  \mathrm{erfc}(\sqrt{\omega})/2$ that we wish to compute through the biased sampling of trajectory endpoints. When $\omega$ is large, diffusion is very slow and the probability $C(L\tau)$ to reach $q_L \ge 1$ is very small. We thus bias the simulation by adding the soft restraint $\mathcal{L}(z)= - 2 \omega q_L$, so as to gradually increase the fraction of trajectories ending in the product basin $q_L \ge 1$ with increasing $\theta$ value. This way of proceeding is similar in spirit to the confining protocol~\eqref{p:confining} which instead is based on a series of hard constraints. 

In this illustration, we represent a path only using its endpoint, denoted by $q$ to simplify. Plugging $\mathcal{A}(\theta) = \ln (\omega/\pi) - \omega \theta^2$ in Eq.~\eqref{eq:condproba}, the conditional probability of $q$ given $\theta$ is equal to 
\begin{equation}
\pi(q | \theta) = \sqrt{\omega/\pi} \exp\left[-\omega(q-\theta)^2 \right] \label{eq:distrib}. 
\end{equation}  
The unbiased distribution is obtained by setting $\theta$ to 0 in~\eqref{eq:distrib} and is represented as a function of $q$ by the green curve delimiting the red area in Fig.~\ref{fig1}. 
The conditional distributions with various biases, obtained by setting $\theta$ to $\tfrac{1}{4}$, $\tfrac{1}{2}$, $\tfrac{3}{4}$ and $1$ in~\eqref{eq:distrib} are displayed. 

Two Brownian motions are considered: a fast one for which $\omega = 5$  and a slow one for which $\omega =100$. The various distributions for the two values of the $\omega$ parameter are displayed in Fig.~\ref{fig1}.a and in Fig.~\ref{fig1}.b, respectively. 
For $\omega=5$, we observe that the biased distributions $q \rightarrow \pi(q|1)$ substantially overlap with both the target region ($q \ge 1$) and the reference distribution $q \rightarrow \pi(q|0)$ (areas displayed in red and green respectively). 
For $\omega=100$, none of the biased distribution ($\theta=\tfrac{1}{4}$, $\tfrac{1}{2}$ and $\tfrac{3}{4}$) substantially overlaps with both the reference distribution and the target region. 
At variance, the marginal distribution~\eqref{eq:pmarg} obtained for $\theta_\mathrm{max}=1$ and both values of $\omega$ (curves displayed in blue) overlap with the unbiased distribution. 

\begin{figure}[!h] \label{fig1}
\includegraphics[width=0.45\textwidth]{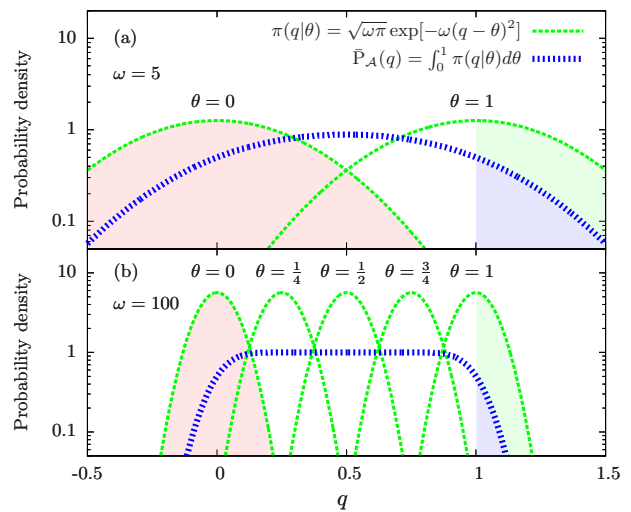}
\protect\caption{Conditional distributions $\pi(q|\theta)$ for the indicated values of $\theta$ (green curves) and the marginal probability of $q$ (blue curves) for $\omega=5$ in pannel (a) and $\omega = 100$ in pannel (b). The target region corresponds to $q\ge 1$ and the unbiased distribution of $q$ is filled in red.}
\label{fig1} 
\end{figure}

A question then naturally arises as to how deviations from our choice of $\mathcal{A}$ as biasing potential qualitatively affects the overlapping properties of the marginal probability of $q$. To answer, we set the biasing potential to $(1+\delta)\mathcal{A}$: the deviation parameter $\delta$ determines how close the biasing potential is to the potential of mean force, the value zero of $\delta$ having been considered so far. We investigate in particular the effect of $\delta$ on the overlapping properties of the maginal probability distributions with respect to the reference distribution and the region of interest. We observe in Fig.~\ref{fig2} that the extent of overlap with the reference distribution increase with increasing values of $\delta$ and that the occurrence of an event concommitantly decreases. Hence, the choice $\delta=0$ offers an interesting qualitative trade-off: the overlap with the reference distribution is substantial and the rare events are frequently observed. 

\begin{figure}[!h]\label{fig2}
\includegraphics[width=0.45\textwidth]{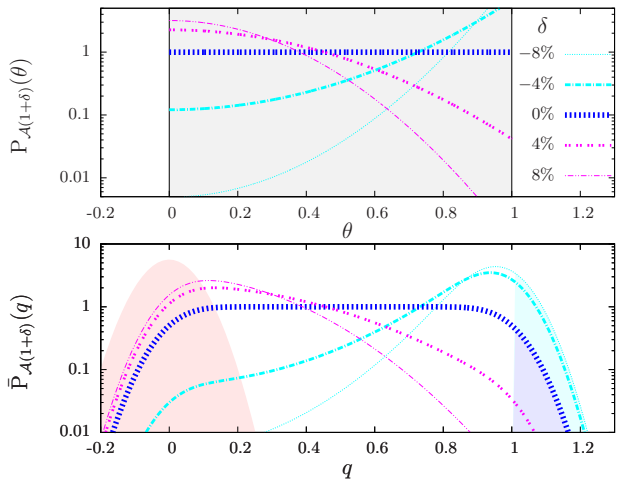}
\protect\caption{ Marginal probabilities of $\theta$ (a) and of $q$ (b) for the biasing potentials $A = (1+\delta) \mathcal{A}$ with varying values of $\delta$.}
\label{fig2} 
\end{figure}

Next, the efficiency of the BF estimator is compared to the one of the corresponding standard estimator (SR). The latter one is derived in Sec.~\ref{sec:testbed}. 
We use the present toy model [Sec.~\ref{subsec:confining}] for computing the $a$-to-$b$ correlation function and investigate the relevant case where the auxiliary biasing potential is set to the potential of mean force. The efficiency of the ABF procedure~\ref{subsec:adaptive} for computing the potential of mean force is demonstrated 
in Sec.~\ref{sec:diffusion} in simulations of vacancy diffusion.

\section{Estimator efficiency~\label{sec:testbed}}

\subsection{Standard reweighting estimator}

The aforementioned method of expanded ensembles~\cite{lyubartsev:1992,iba:2001} consists of sampling the extended space $(\vartheta,z) \in \Theta \cup \mathcal{Z} $ equipped with joint probability 
\begin{equation} \label{jointproba}
 \mathrm{p}_{A} (\vartheta, z ) \propto h_a(q_0) \exp \left[  A (\vartheta) - \beta \mathcal{H} (z) - \mathcal{K}(\vartheta,z)\right] 
\end{equation}
which is formally equal to $\pi(z|\vartheta) \mathrm{P}_{A} (\vartheta)$. This sampling approach alternatively suggests using the standard reweighting estimator in order to correct for the sampling biases. However, the SR estimator does not converge well in the present rare-event context, whether the biased sample be generated according to the joint probability distribution~\eqref{jointproba} or to the restricted distribution $\pi(z|\theta_{\mathrm{max}})$, wherein $\vartheta$ is set to $\theta_\mathrm{max}$. This limitation is illustrated in Sec.~\ref{subsec:comparison} and~\ref{subsec:fluctuation} by performing simulations in the expanded ensemble so as to allow direct comparison with the (more efficient) BF estimator presented in Sec.~\ref{subsec:BF_estimator}. The SR estimator associated with observable $\mathcal{O}$ is constructed from the following expectation ratio  
\begin{equation}
 \mathbb{E}_\pi (\mathcal{O} |\theta ) = \frac{  c_\Theta \displaystyle \int \limits_{\Theta\cup\mathcal{Z}} \mathcal{O}(z) \dfrac{\mathrm{p}_A(\theta, z)}{\mathrm{p}_A(\vartheta,z)} \mathrm{p}_A (d\vartheta,\mathcal{D}z) }
 { c_\Theta \displaystyle \int\limits_{\Theta\cup\mathcal{Z}} \dfrac{ \mathrm{p}_A(\theta, z) }{ \mathrm{p}_A(\vartheta,z)} \mathrm{p}_A (d\vartheta,\mathcal{D}z)} \label{eq:expectation_std}. 
\end{equation}
The normalizing constant $c_\Theta = 1/\left(\int_\Theta d\vartheta\right)$ is added to the numerator and denominator of ratio~\eqref{eq:expectation_std} so that the latter quantity is formally equal to $\mathrm{P}_A(\theta)$, the marginal probability of $\theta$. Resorting to the ergodic theorem, the conditional expectation of observable $\mathcal{O}$ given $\theta$ and $\left\{ \vartheta^m , z^m \right\}_{1 \leq m \leq M}$, a Markov chain constructed using a Monte Carlo scheme leaving the joint probability distribution~\eqref{jointproba} invariant, may be estimated using (SR estimator) 
\begin{equation} \label{eq:estimator_std}
 \hat{\mathbb{E}}^{\mathrm{S},M}_\pi \left( \mathcal{O} | \theta \right) = 
 \frac{\tfrac{c_\Theta}{M} \sum\limits_{m=1}^M  \mathcal{O}^m e^{ A(\theta) -\mathcal{K}(\theta,z^m) }  \left/ e^{ A(\vartheta^m) - \mathcal{K}(\vartheta^m,z^m )} \right. } {\tfrac{c_\Theta}{M} \sum\limits_{m=1}^M  e^{ A(\theta)-\mathcal{K}(\theta,z^m)} \left/ e^{ A(\vartheta^m) - \mathcal{K}(\vartheta^m,z^m)} \right.}. 
\end{equation} 
Remarkably, the BF estimator~\eqref{eq:estimator_std} can also be implemented when the sample is generated according to the joint distribution. To justify its use, one simply resorts to the ergodic theorem within the following expectation ratio: 
\begin{eqnarray}
 \mathbb{E}_\pi \left( \mathcal{O} | \theta \right)   & = & \frac{\int_{\Theta \cup \mathcal{Z}} \mathcal{O}(z) \bar{\pi}_A (\theta | z )  \mathrm{p}_A (d\vartheta ,\mathcal{D} z) } {\int_{\Theta \cup \mathcal{Z}} \bar{\pi}_A (\theta|z)  \mathrm{p}_A (d\vartheta ,\mathcal{D} z)} \label{eq:ratio_b}
\end{eqnarray}
Relation~\eqref{eq:ratio_b} is a consequence of Bayes formula as it is obtained by plugging $\bar{\mathrm{P}}_{A} ({z}) = \int_{\Theta} \mathrm{p}_{A} (d \vartheta, {z} )$ into~\eqref{eq:ratio_a}. 

Estimates associated with the unbiased path distribution are recovered by setting $\theta$ to 0 in~\eqref{eq:estimator}. The BF estimator simplifies to
\begin{equation} \label{eq:BF}
 \widehat{\mathcal{O}}^{\mathrm{B},M} = \frac{\tfrac{1}{M}\sum\limits_{m=1}^M \mathcal{O}^m e^{ A(0)}\left/ \left(  \int\limits_\Theta e^{ A(\vartheta) - \mathcal{K}(\vartheta,z^m)} d \vartheta \right) \right. } {\tfrac{1}{M}\sum\limits_{m=1}^M  e^{ A(0)} \left/\left( \int\limits_\Theta e^{ A(\vartheta) - \mathcal{K}(\vartheta,z^m) } d \vartheta \right) \right.}
\end{equation}
Resorting to the sequence of biasing potentials $ B^m = \ln \int_\Theta  \exp \left[ A (\vartheta) -  \mathcal{K}(\vartheta,z^m) \right]  d\vartheta \label{bias} $, we may write estimator in the conventional form~\eqref{eq:conventional} mentioned in the introduction. The SR estimator with respect to the unbiased path distribution ($\theta =0$) reads
\begin{eqnarray} \label{eq:SR}
\widehat{\mathcal{O}}^{\mathrm{S},M} & = & \frac{\tfrac{c_\Theta}{M}\sum_{m=1}^M \mathcal{O}^m e^{ A(0)-A(\vartheta^m) + \mathcal{K}(\vartheta^m ,z^m) } } { \tfrac{c_\Theta}{M}\sum_{m=1}^M e^{ A(0)-A(\vartheta^m) + \mathcal{K}(\vartheta^m,z^m)}}. 
\end{eqnarray}
The denominators in Eqn.~\eqref{eq:BF} and~\eqref{eq:SR} both yield estimates of $\mathrm{P}_A(\theta)$ with $\theta=0$. 
Relevant information about the estimator convergence can be learned by monitoring the marginal probability of $\theta$. 

Note that $A(\vartheta^m) - \mathcal{K}(\vartheta^m,\mathcal{L}^m)$ corresponds to the biasing potential associated with path $z^m$. 
Compared with the SR estimator~\eqref{eq:SR}, the sampled values of the external parameter $\theta$ are irrelevant information in the BF estimator~\eqref{eq:BF}. 
We now show that these  distinct features results in different numerical efficiencies. 

\subsection{Comparison of BF and SR estimators in a generic model~\label{subsec:comparison}}

We consider the toy model of Sec.~\ref{subsec:confining} in which the biasing potential has already converged to the potential of mean force and $\theta_{\mathrm{max}}=1$, implying that the marginal probability density of $\theta$ is equal to one in $\Theta$ and that the conditional probability of $q$ given $\theta$ in~\eqref{eq:distrib} is equal to the joint probability $\pi(q|\theta) = \sqrt{\omega/\pi} \exp\left(-\omega(q-\theta)^2 \right)$. 

The correlation function $C(L\tau)$ is estimated using BF estimator~\eqref{eq:BF} and SR estimator~\eqref{eq:SR} by setting the observable $\mathcal{O}$ to $h_b(q)$ equal to $1 $ if $q\ge 1$ and to $0$ otherwise. The marginal probability of $\theta$ at 0 is also estimated from the denominator of~\eqref{eq:BF} and~\eqref{eq:SR}. 
Each generated Markov chain $\left\{ \vartheta^m, q^m \right\}_{1 \leq m \leq M}$ is used twice, first to obtain a BF estimate based on~\eqref{eq:BF} and then to get a SR estimate based on~\eqref{eq:SR}. The $m$th state of the chain is generated as follows: $\vartheta^m$ is drawn randomly and uniformly in $\Theta=[0,1]$ interval and $q^m$ is drawn in the Gaussian distribution of $(2\omega)^{-1}$ variance and $\vartheta^m$ mean. Displayed in Fig.~\eqref{fig3} are the means and standard errors of $10^5$ independent estimates, obtained using both estimators. 
We observe that, with increasing $\omega$ parameter, only the BF estimator yields an accurate estimation of the marginal probability of $\theta$ at $0$ (Fig.~\ref{fig3}.a) and of the correlation function (\ref{fig3}.b). The computational speed-up of convergence that is achieved by using BF estimator rather than SR estimator can be assessed from their respective standard errors plotted as a function of $\omega$ in Fig.~\ref{fig3}.c. As soon as $\omega$ becomes larger than 20, the standard error associated with BF estimator is two orders of magnitude lower than the one obtained using the SR estimator. The use of Bayes formula accelerates the simulations by about four orders of magnitude. 

\begin{figure}[!h]\label{fig3}
\includegraphics{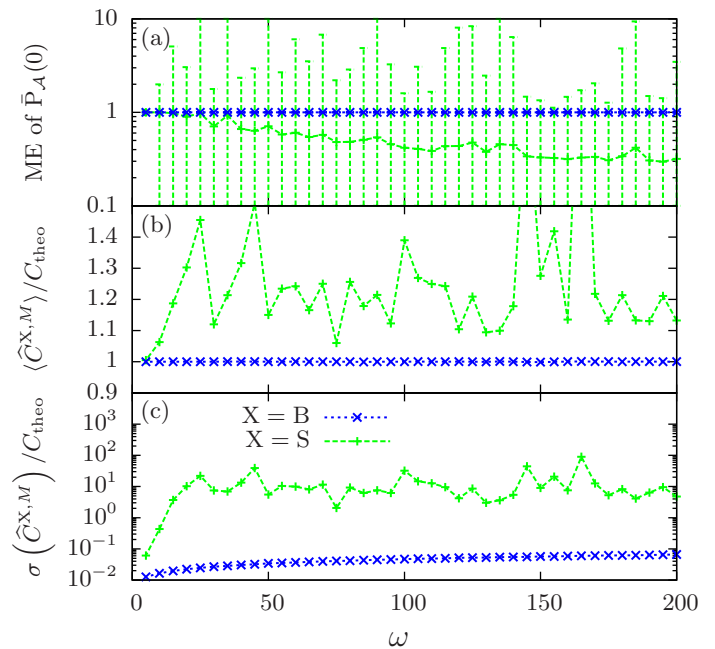}
\protect\caption{Comparison between the Bayesian~\eqref{eq:estimator} and standard~\eqref{eq:SR} estimators: (a) averaged estimates of the marginal probability of $\theta$ at $0$ as a function of $\omega$, (b) averaged estimates of the probability $C(M\tau)$ normalized to the exact probability and (c) its normalized standard error.  All averages are obtained from $10^5$ estimates and each estimate is obtained using $M=10^4$ points. }
\label{fig3} 
\end{figure}

\subsection{Fluctuation relations \label{subsec:fluctuation}}

\begin{figure}[!h]
\includegraphics[width=0.45\textwidth]{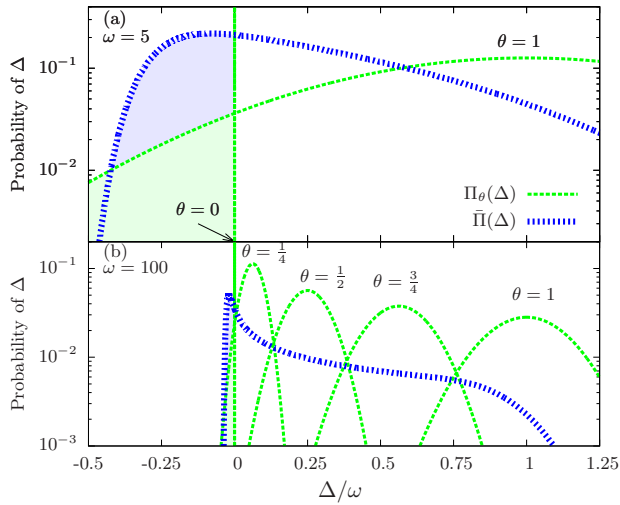}
\protect\caption{Probability distribution of $\Delta$ as a function of the $\Delta$, blue curve for BF estimator and green curves for SR estimator with various values of $\theta$.}
\label{fig4}
\end{figure}

To explain why BF and SR estimators behave differently, we resort to fluctuation relations~\cite{bennett:1976,athenes:2007} in order to analyse how the quantity $\mathrm{P}_\mathcal{A}(0)$ is evaluated in both approaches. This probability  corresponds to the denominator of ratio~\eqref{eq:BF} for the BF estimator, and of ratio~\eqref{eq:SR} for the SR estimator, when $A$ is set to $\mathcal{A}$. The probability density is one when $\Theta$ is the $[0,1]$ interval.  This probability is computed from two different ensemble averages, entailing two distinct types of fluctuation relations. For the BF approach, the following single relation must hold 
\begin{equation}
 \langle e^{-\Delta(q)} \rangle = \int_\mathbb{R} e^{-\Delta} \bar{\Pi}(d \Delta) = e^0 \label{eq:fluctuation}
\end{equation}
where $ \Delta(q) = - \ln \left[ \mathrm{P}_\mathcal{A}(0,q)/\bar{\mathrm{P}}_\mathcal{A}(q)\right]$ and $\bar{\Pi}(\Delta)$ denotes the probability that $\Delta(q)$ takes the particular value $\Delta$. As for the SR approach, a whole family of detailed fluctuation relations must be satisfied by the quantity $ \Delta(\vartheta,q) = - \ln \left[ \mathrm{P}_\mathcal{A}(0,q)/\mathrm{P}_\mathcal{A}(\vartheta,q)\right]$ 
\begin{eqnarray}
 \langle e^{-\Delta(\theta,q) }  \rangle_\theta & = & \int_{\mathbb{R}} e^{-\Delta} \Pi_\theta (d\Delta) = e^0 \label{eq:fluctuation2} , \\
 \langle e^{-\Delta(\vartheta,q)}  \rangle      & = & \int_\Theta   \langle e^{-\Delta (\theta,q)} \rangle_\theta \mathrm{P}_\mathcal{A} (d\theta) =e^0 \label{eq:overallfluctuation}
 \end{eqnarray}
where $\Pi_\theta (\Delta)$ denotes the probability that $\Delta(\theta,q)$ takes the particular value $\Delta$. The uniform average over $\theta \in \Theta$ yields an additional overall fluctuation relation~\eqref{eq:overallfluctuation} corresponding to~\eqref{eq:fluctuation}. 
Because the exponential function is strictly increasing, the negative values of $\Delta$ and the stricly positive ones have similar statistical weight in~\eqref{eq:fluctuation} or~\eqref{eq:fluctuation2}, in the sense that the sum of the two contibutions are equal to one (except for the case $\theta = 0$). 
Hence, excessively small fractions of negative $\Delta$ values will result in large statistical variance and in slow numerical convergence of the estimates as a function of simulation time.~\cite{bennett:1976} The probabilities  $\Pi_\theta (\Delta)$ with $\theta \in \left\{ 0, \tfrac{1}{4}, \tfrac{1}{2}, \tfrac{3}{4}, 1\right\}$ and $\bar{\Pi}(\Delta)$ have been plotted as a function of $\Delta$ for $\omega=5$ in Fig.~\ref{fig4}.a and and for $\omega=100$ in Fig.~\ref{fig4}.b. From the distributions at $\omega=100$, we observe that the probability to have  $\Delta(\vartheta^m,q^m) \leq 0$ is negligible when $\vartheta^m > \tfrac{1}{2}$. Given the fact that the $\vartheta^m$'s are sampled uniformly in $[0,1]$, substantial deviations from fluctuation relations~\eqref{eq:fluctuation2} will inevitably be measured in typical (finite-length) simulations, resulting in inaccurate SR estimates. At variance, the BF approach does not suffer from this limitation, as the fluctuation relation that must be satisfied is 
global. We indeed observe that the fraction of the negative values of $\Delta$ is always substantial, making the BF estimator particularly efficient for large $\omega$ values. 

\section{Application to vacancy migration in $\alpha$-Fe~\label{sec:diffusion}}

We now demonstrate the efficiency of the approach in simulations of the migration of a single vacancy on a lattice in $\alpha$-Fe, a crystalline phase of iron with body-centered cubic (BCC) structure. The migration corresponds to the jump of an atom in the [111] direction into a nearest neighbor vacant site. 
The length of the jump is $\mathrm{a}_0 \sqrt{3}/2$, where $\mathrm{a}_0$ is the side of the BCC cube. 
Atomic interactions of this atomic system are described by an embedded atom model potential.~\cite{ackland2004} 
Reference values for migration rates are available for this testbed system that was previously investigated in Ref.~\onlinecite{athenes:2012} using a combination of transition path sampling and the multistate Bennett acceptance ratio method for postprocessing. 

\subsection{Computational set-up}

The computational set-up is as follows. Basin $a$ and $b$ are defined with respect to the underlying perfect lattice whose sites are the atomic positions of the structure at 0 K without the vacancy. The lattice parameter is $\mathrm{a}_0=2.8553 \textrm{\AA}$. 
The indicator function $h_a$ is equal to 1 if all atoms are located within a distance of $0.45 \textrm{\AA}$ from their lattice site, and to 1 otherwise. 
The characteristic function $h_b$ is 1 if one atom is located beyond a distance of $\mathrm{a}_0\sqrt{3}/4$ from its lattice site, otherwise it is 0. 
Path sampling consists of shooting and shifting moves as detailed in the Appendices~\ref{shooting} and~\ref{shifting}. 
Trajectories contain $L=150$ steps with time-step $\tau=2fs$. A position-Verlet scheme~\cite{athenes:2012} is used to construct $x_{\ell+1}$ from $x_\ell$, meaning that the gradient of the potential energy is evaluated at $q_{\ell+1/2}=q_{\ell}+p_\ell \tau/2$. The Jacobian matrix associated with the MD transformation exhibits eigenvalues that are either complex numbers located on the unit circle or real positive numbers. Let us denote by $\mu_{\ell+1/2}$ the logarithm of the smallest eigenvalue modulus. 
Its value is characterized by the eigenvalue spectrum of the Hessian matrix associated with the potential energy at $q_{\ell+1/2}$.
Details on the connection between the Hessian and Jacobian matrices are given in Ref.~\onlinecite{athenes:2012}. 
The value of $\mu_{\ell+1/2}$ is strictly negative when the lowest egeinvalue of the Hessian matrix is strictly negative, in which case the energy surface is negatively curved along the direction generated by the corresponding eigenvector. 
The biasing path functional is set to 
\begin{equation}
 \mathcal{L}(z) = \max \left[ \mathcal{L}_{\mathrm{min}} , \sum_{\ell=0}^{L-1} \mu_{\ell+1/2} \right] \leq 0
\end{equation}
The cut-off parameter $\mathcal{L}_{\mathrm{min}}$ is set to the value $-9$. It is used to prevent from exploring regions containing second order saddles and thus to save computational time,  trajectories leading to such regions corresponding to non-reactive rare events. The lowest eigenvalues of the Hessian is computed using the Lanczos algorithm.~\cite{lanczos1961applied}
Details about the numerical implementation are given in Ref.~\onlinecite{cances2009,marinica2011energy}. 

The illustrations are given at the temperature of $500$ K. We set $\theta_\mathrm{max}=2.1$. In the following, a simulation run utilizes $K=480$ replicas and consists of $M=10^4$ Monte Carlo cycles. Each replica is allocated to a distinct processor. A cycle consists of performing a shooting move followed by a shifting move for each replica. 
The two procedures are detailed in Appendices~\ref{shooting} and~\ref{shifting}, as their implementation slightly differs from the one given previoulsy~\cite{athenes:2012}. 
\subsection{Construction of the auxiliary potential}

\begin{figure}[!h]
\includegraphics[width=0.45\textwidth]{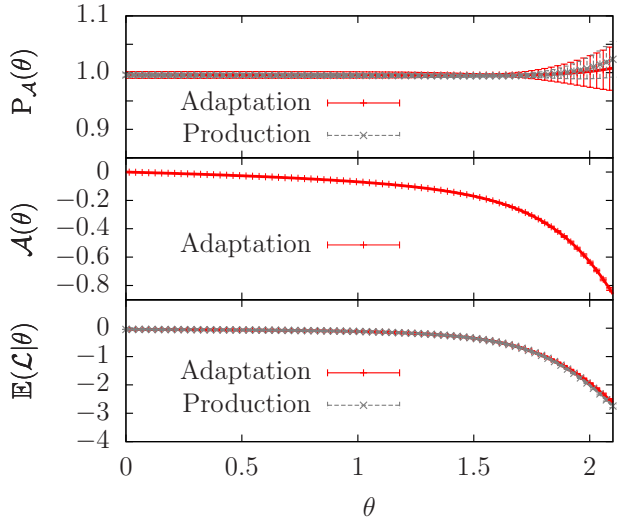}
\protect\caption{Marginal probability of $\theta$ (multiplied by $\theta_\mathrm{max}$), potential of mean force and mean force as a function of $\theta$.}
\label{fig5}
\end{figure}

Two consecutive series of 5 independent simulation runs are performed. The first five runs aim at constructing the auxiliary biasing potential using the ABF scheme~\eqref{eq:abf_scheme}. Then, freezing the previoulsy obtained biasing potentials, 5 subsequent (production) runs are performed to estimate the expectations using the scheme~\eqref{eq:estimator_K}. Figure~\ref{fig5} displays the estimates of $\mathrm{P}_\mathcal{A}(\theta)$, $\mathcal{A}(\theta)$ and $\mathbb{E}(\mathcal{L}|\theta)$ as a function of $\theta$ and averaged over the 5 runs. The standard errors are evaluated from the 5 estimates and are indicated by error bars for the three quantities in  Fig.~\ref{fig5}. We observe that a flat histogram is obtained for the marginal probability of $\theta$. Reproductible data are obtained for the mean force and its potential. Furthermore, the difference between the adaptive and production runs is insignificant, final averages could have been taken after the adaptation run. 

\begin{figure}[!h]
\includegraphics[width=0.45\textwidth]{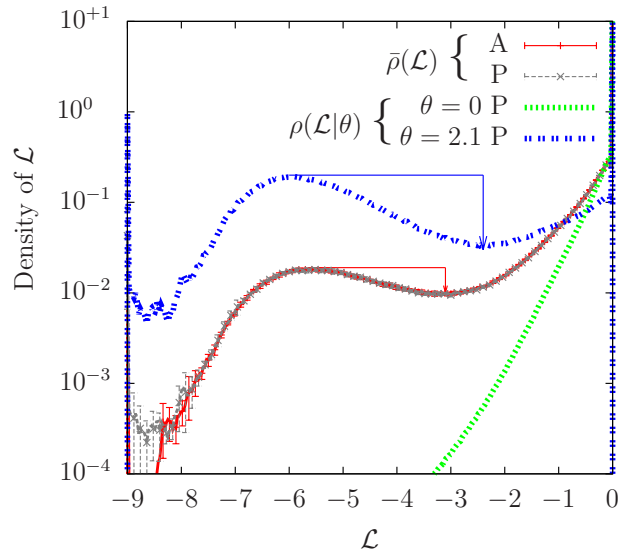}
\protect\caption{Distributions of $\mathcal{L}$: $\bar{\rho}( \mathcal{L})$ and $\rho( \mathcal{L} |\theta)$ denote the probabilities that $\mathcal{L}(z)$ takes value $\mathcal{L}$ with respect to path distributions $\bar{\mathrm{P}}_{\mathcal{A}}(z)$ and $\pi(z|\theta)$, respectively.  }
\label{fig6}
\end{figure}

The standard errors are small and not clearly visible on the graphs in Fig.~\ref{fig5}, except on the curve displaying the marginal probability of $\theta$. 
We observe that the standard errors associated with $\mathrm{P}_\mathcal{A}(\theta)$ increase with $\theta$ and become substantial at large $\theta$ values. To explain this trend, let us examine $\bar{\rho}( \mathcal{L})$, the probability distributions of the sampled $\mathcal{L}$ values. We observe in Fig.~\ref{fig6} that this distribution is bimodal. The large peak at 0 corresponds to typical trajectories that are non reactive. The smaller peak in the range from $-7$ to $-4$ contains both reactive trajectories and active trajectories returning to $a$. The presence of two peaks means that metastability is not completely suppressed through path-sampling, even though the fraction of reactive trajectories is enhanced by several orders of magnitude compared with the one associated with the unbiased distribution $\rho(\mathcal{L}|0)$. Concerning the biased distribution $\rho(\mathcal{L}|\theta_{\mathrm{max}})$, the peak containing the reactive trajectories is higher and more pronounced than that of the sampled 
distribution. This feature explaining the substantial statistical fluctuations observed in the measurement of $\mathrm{P}_{\mathcal{A}}(\theta)$ when $\theta$ is large. 
Note that the $\theta_c$ value for which the two peaks of the bimodal distribution $\rho(\mathcal{L}| \theta_c)$ have equal weights occurs in the range $2.2\--2.4$ and decreases with the path length~\cite{athenes:2012}. Here, $\theta_c$ would correspond to the inflexion of the $\mathbb{E}(\mathcal{L}|\theta)$ curve, outside the plot in Fig.~\ref{fig5}.c. As reported in Ref.~\onlinecite{athenes:2012}, the restricted sampling of the conditional distribution $\pi(z|\theta)$ becomes very difficult when $\theta > \theta_c$, the measured autocorrelation function of $\mathcal{L}$ increases drastically. Important autocorrelations are also observed in the sampling of $\bar{\mathrm{P}}_{\mathcal{A}}(z)$ distribution when $\theta_{\mathrm{max}}$ is set to a value larger than $\theta_c$. However, the extent of metastability is smaller with the approach based on Bayes formula. We speculate that this results from the smaller barrier height for trajectory disactivation for $\bar{\rho}(\mathcal{L})$ distribution than for $\
rho(\mathcal{L}|\theta)$ distribution, as indicated in Fig.~\ref{fig6} by the red and blue downward arrows, respectively.  

\begin{figure}[!h]
\includegraphics[width=0.45\textwidth]{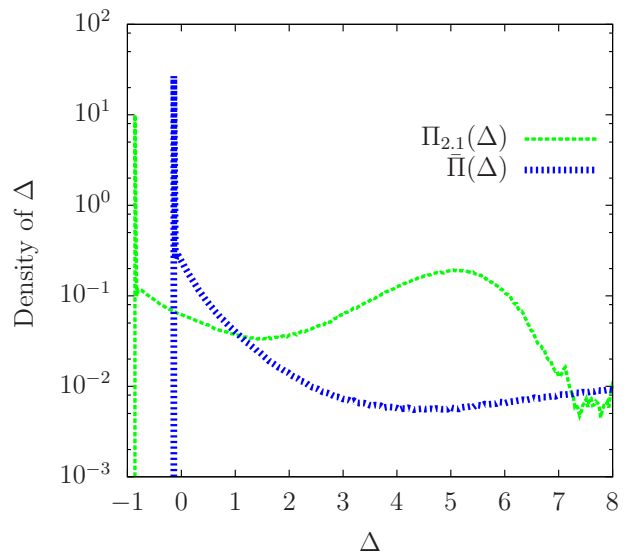}
\protect\caption{Distribution of $\Delta$ defined in Sec.~\ref{subsec:fluctuation}. }
\label{fig7}
\end{figure}

Another advantage of the approach based on Bayes formula is illustrated from the fluctuation theorems defined in Sec.~\ref{subsec:fluctuation}. We observe in Fig.~\ref{fig7} that the distributions of the $\Delta$ values are more peaked around $0$ when the marginal probability is sampled, a result expected and consistent with the previous analysis using the toy model. Numerical convergence should therefore be faster for the reason mentioned in Sec.~\ref{sec:testbed}

\begin{figure}[!h]
\includegraphics[width=0.45\textwidth]{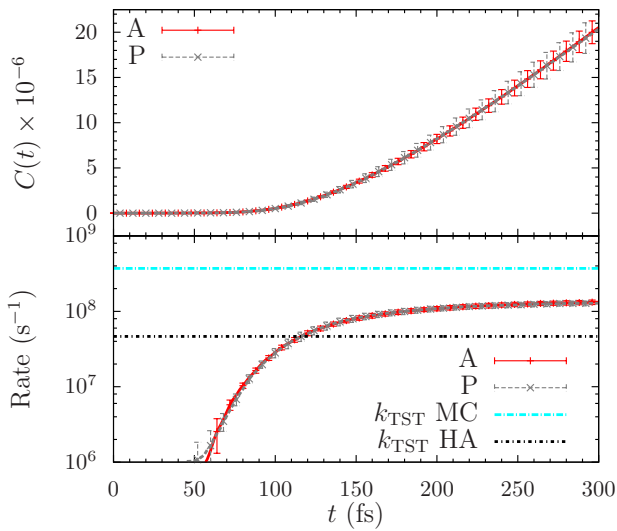}
\protect\caption{Time correlation function and its derivative as a function of the time $t=\ell \tau$, where $\ell$ is the index of the trajectory states. The horizontal lines corresponds to the values obtained using transition state theory~\cite{HTB1990,athenes:2012} in which the free energy barrier is evaluated using either Monte Carlo simulations (MC) or the classical harmonic approximation (HA).}
\label{fig8}
\end{figure}

The sampled path distributions with $\theta_{\mathrm{max}} < \theta_c$ contain high enough a fraction of reactive trajectories so as to accurately estimate the $a$-to-$b$ correlation function. 
Figure~\ref{fig8} represents the time correlation function and its time derivative as obtained after the 5 adaptation runs and the 5 production runs. 
Smooth values are obtained for the time-derivative owing to recycling of the shifted trajectories using the waste-recycling procedure described in Appendix~\ref{waste-recycling}. 
The phenomenological transition rate corresponds to the plateau value, which is in perfect agreement with the value previousy calculated in Ref.~\onlinecite{athenes:2012}. 

The present simulations involving the hopping of a vacancy in a crystal show that the approach can easily be implemented on a parallel computer architecture by propagating several replicas of the system simultaneously and adapting the biasing force periodically. The proposed framework allows to explore the multiple reaction channels corresponding to the 8 possible atomic jumps into the vacancy. This is a clear advantage compared with other rare event approaches such as transition interface sampling,~\cite{vanerp:2003} and forward flux sampling~\cite{allen:2006} that tends to confine trajectories into separate transition channels. The use of Bayes formula in TPS method thus facilitates its implementation and will certainly extend its scope. 

\section{Conclusion}
 
Estimating the state-to-state correlation functions in TPS method requires to extract the free energy difference associated with the transformation of a trajectory ensemble into a perturbed ensemble wherein the rare reactive trajectories have become frequent. This task had been achieved so far through postprocessing of the information contained in a series of Markov chains, in which the strength of the perturbating bias favouring the occurrence of reactive trajectories was gradually increased. TPS was therefore a most relevant case study to apply the recently proposed approach~\cite{cao:2014} for computing free energies based on Bayes formula and adaptive biasing. 
The combination of the two techniques allows to adaptively construct a biased sample whose associated distribution substantially overlaps with both the unpertubed and perturbed distributions and to \emph{in fine} obtain an unbiased estimate of the time-correlation function. 

The ability of the proposed approach to construct and sample a biased distribution that substantially overlaps with any pertubed and unperturbed distributions is extremely useful in general. From a larger perspective, this feature will enable one to automatically remediate the convergence issue that is encountered in umbrella sampling when the perturbed distribution, sampled by this FEP-based simulation method, insufficiently overlaps with the reference distribution. This last situation is quite ubiquitous in molecular simulation and, not incidentally, motivated the testbed model investigation of Sec.~\ref{sec:testbed}. We thus expect many interesting applications of the Bayes formula approach in order to compute rare-event frequencies or mean-force potentials in domains ranging from chemistry to bio-physics and materials science. 

\begin{acknowledgments}
Stimulating discussions with G. Stoltz are acknowledged. This work was performed using HPC resources from GENCI-[CCRT/CINES] (Grant x2015096973). 
\end{acknowledgments}

\bibliographystyle{prl}

\appendix

\section{Shooting moves\label{shooting}}

A shooting move consists in performing the following operations ($z$ is current path)
\begin{enumerate}[i.]
 \item draw an integer $\ell$ randomly and uniformly in $\{ 0, 1, \cdots, L \}$ 
 \item from $x_\ell \in z $, generate a neighboring state $\tilde{x}_\ell$ according to the canonical distribution \label{tube} ;
 \item construct the trial trajectory $\tilde{z}$ by applying the Verlet map $L-\ell$ times forward from $\tilde{x}_\ell$ and $\ell$ time backward from $\tilde{x}_\ell$; 
 \item compute the new path biasing potential 
 \begin{equation}
 B(\tilde{z}) = \ln \int_\Theta  \exp \left[ A (\theta) - \mathcal{K}(\theta,\tilde{z}) \right]  d\theta \label{bias}
\end{equation}
 \item draw a random number $R$ uniformly in $(0,1]$; if $R$ is lower than the acceptance probability
 \begin{equation}
  \mathrm{P_{acc}} [\tilde{z} \leftarrow z] = \min \left\{ 1, h_a(\tilde{q}_0) \exp \left[B(\tilde{z}) - B(z) \right] \right\}, \label{metro}
 \end{equation}
 accept the trial trajectory, otherwise reject it. 
\end{enumerate}

This scheme is correct because the probability to accept a trial path in~\eqref{metro} corresponds to a proper Metropolis-hasting acceptance probability. We have   
\begin{equation}
  \mathrm{P_{acc}} [\tilde{z} \leftarrow z] = \min \left\{ 1, \frac{\mathrm{P_{gen}} [z \leftarrow \tilde{z}] \bar{\mathrm{P}}_A(\tilde{z}) }{\mathrm{P_{gen}} [\tilde{z} \leftarrow z] \bar{\mathrm{P}}_A(z)} \right\}, \label{metroformal}
 \end{equation}
 where $ \mathrm{P_{gen}} [\tilde{z} \leftarrow z] $ is the probability to generate $\tilde{z}$ from $z$ and vice versa for $\mathrm{P_{gen}} [z \leftarrow \tilde{z}]$. In practice, the path generating procedure in~\ref{tube} constructs new momenta $\tilde{p}_\ell$ in the canonical distribution by slightly perturbating the momenta $p_\ell$ using an Ornstein-Uhlenbeck process~\cite{stoltz:2007,athenes:2012}. We use $\tilde{p}_\ell = \epsilon p_\ell + \sqrt{1-\epsilon^2} \varpi$ where $\varpi$ are momenta drawn in the Maxwell-Boltmzann distribution at temperature $\beta^{-1}$. The parameter $\epsilon \in [0,1]$ mixes the uncorrelated momenta $p_\ell$ and $\varpi$. 
 As a result, the following condition is satisfied 
\begin{equation}
 \frac{\mathrm{P_{gen}} [z \leftarrow \tilde{z}] }{\mathrm{P_{gen}} [\tilde{z} \leftarrow z]}  = \frac{\exp [-\beta \mathcal{H}(\tilde{z})]}{ \exp [-\beta \mathcal{H}(z)]}. \label{ratio}
\end{equation}
Noticing that the marginal probabilities of path $z$ and $\tilde{z}$ in~\eqref{metroformal} are proportional to $h_a(q_0) e^{B(z)-\beta \mathcal{H}(z)}$ and $h_a(\tilde{q}_0) e^{B(\tilde{z})-\beta \mathcal{H}(\tilde{z})}$ respectively, and plugging~\eqref{ratio} into the formal rate~\eqref{metroformal}, the simple form~\eqref{metro} is obtained. It is then easy to show that detailed balance with respect to the sampled distribution is obeyed.~\cite{frenkel:2002,lelievre:2010} 
In the numerical application given in Sec.~\ref{sec:diffusion}, the value of the mixing parameter is $\epsilon=0.975$, which yields a mean acceptance rate of $90.6\%$. 

\section{Shifting moves\label{shifting}}

Let $\bar{z} =  \left\{\bar{x}_{h} \right\}_{0 \leq h \leq 2L }$ denote a path of $2L$ states generated by the Verlet map and $\tilde{z}_\ell = \left\{\bar{x}_{h+\ell} \right\}_{0 \leq h \leq L}$ with $0\leq \ell \leq L$ denote the $L+1$ path segments of $L$ states included in $\bar{z}$. The path space for the extended paths is $\bar{\mathcal{Z}}$. We define $\bar{\mathrm{Q}} (\bar{z} | z )$ the conditional probability to generate $\bar{z}$ from $z$ using the Verlet map. We have   
\begin{equation}
 \bar{\mathrm{Q}} (\bar{z} | z ) = \begin{cases}
					  \frac{1}{L+1} & \text{if } \exists \ell \in \left\{0, 1, \cdots, L \right\} :  \tilde{z}_\ell = z  \\
					  0 & \text{otherwise}. 
                                   \end{cases}
\end{equation}
The form of $\mathrm{Q}(\bar{z}|z)$ reflects the fact that there are exactly $1+L$ distinct paths $\bar{z} \in \bar{\mathcal{Z}}$ (generated through $2L$ successive applications of the Verlet map) such that $\bar{z}$ includes path $z \in \mathcal{Z}$. 
The marginal probability of $\bar{z}$ then follows 
\begin{eqnarray}
 \bar{\mathcal{P}}_A (\bar{z}) & = & \int_{\mathcal{Z}} \bar{\mathrm{Q}} (\bar{z} | z ) \bar{\mathrm{P}}_A (z ) \mathcal{D} z \\
                               & = & \frac{1}{L+1}\sum_{\ell =0 }^{L} \bar{\mathrm{P}}_A (\tilde{z}_\ell) 
\end{eqnarray}
where $z \in \mathcal{Z}$. We also define the probability to select $\bar{z}_\ell$ from $\bar{z}$ in the shifting procedure as 
\begin{eqnarray}
 \mathrm{P}_{\mathrm{sel}}(\ell | \bar{z}) = \frac{ \mathrm{Q}(\bar{z}| \tilde{z}_\ell ) \bar{\mathrm{P}}_A (\tilde{z}_\ell) }{ \bar{\mathcal{P}}_A(\bar{z}) } & = & \frac{ h_a(\bar{x}_\ell) \exp \left[ B(\tilde{z}_\ell) \right] }{ \sum_{l = 0}^{L} h_a(\bar{x}_l) \exp \left[ B(\tilde{z}_l ) \right]}. 
\end{eqnarray}

A shifting move then consists in performing the following operations ($z$ is current path)
\begin{enumerate}[i.]
\item draw an integer $\ell$ randomly and uniformly in $\{ 0, 1, \cdots, L \}$ ;
\item set $\bar{x}_{h} = x_{h-\ell}$ for $\ell \leq h \leq L+\ell$ ;
\item construct $\{ \bar{x}_h \}_{L+\ell < h \leq 2L }$ by applying the Verlet map $L-\ell$ times forward from $\bar{x}_{L+\ell}$;
\item construct $\{ \bar{x}_h \}_{0 \leq h < \ell }$ by applying the Verlet map $\ell$ times backard from $\bar{x}_{\ell}$;
\item set $\bar{z} = \left\{\bar{x}_{h} \right\}_{0 \leq h \leq 2L}$;
\item select path $\tilde{z}_\sigma$ by drawing $\sigma$ in the multinomial distribution of probability vector $\sigma \rightarrow \mathrm{P}_{\mathrm{sel}}(\sigma | \bar{z})$;
\item update the state indexes of new path $z= \left\{ x_h \right\}_{0 \leq h \leq L}$ by setting $x_h = \bar{x}_{\sigma + h }$ for $0 \leq h \leq L$;
\end{enumerate}

The proof that shifting moves leave the probability distribution $\bar{\mathrm{P}}_A(z)$ invariant follows the same lines as those given in Ref.~\onlinecite{athenes:2012}. 
It can additionally be shown that the marginal probability $\bar{\mathcal{P}}_A(\tilde{z})$ is also left invariant by the samping procedures. 
This property is useful as it allows to estimate the observable using a waste-recycling estimator detailed hereafter. 

\section{Waste-recycling estimator\label{waste-recycling}}
Using the shifting proposal probability defined in Appendix~\ref{shifting}, we express the conditional probability of $(\theta, z)$ given $\bar{z}$ using the following Bayes formula  
\begin{eqnarray}\label{eq:BFWR}
 \mu_A (\theta , z | \bar{z})  & = & \frac{\mathrm{Q}(\bar{z}|z) \mathrm{p}_A (\theta  , z) }{ \bar{\mathcal{P}}_A (\bar{z}) }, \\
                                    & = &  \pi_{A}(\theta|z) \times \frac{\mathrm{Q}(\bar{z}|z) \bar{\mathrm{P}}_A (z) }{ \bar{\mathcal{P}}_A (\bar{z}) }   . 
\end{eqnarray}
When $z$ is equal to  $\tilde{z}_\ell$ ($\tilde{z}_\ell \in \bar{z}$), the conditional probability simplifies into
\begin{eqnarray}
 \mu_A (\theta , \ell | \bar{z})    & = & \pi_A ( \theta | \tilde{z}_\ell ) \mathrm{P}_{\mathrm{sel}}(\ell | \bar{z}) . 
\end{eqnarray}
Denoting $\mathcal{O} (\tilde{z}^{km}_\ell)$ by $\mathcal{O}^{km}_\ell$, the waste-recycling estimator at $\theta=0$ is based on Bayes formula~\eqref{eq:BFWR} and writes  
\begin{equation} \label{eq:WR}
 \widehat{\mathcal{O}}^{\mathrm{W},K,M} = \frac{\sum_{k=1}^K\sum_{m=1}^M \sum_{\ell = 0}^L \mathcal{O}^{km}_\ell \mu_A (0 ,\ell |  \tilde{z}^{km})} {\sum_{k=1}^K \sum_{m=1}^M \sum_{\ell = 0}^L \mu_A (0 , \ell | \tilde{z}^{km}) }. 
\end{equation}

The waste-recycling estimator can be used to estimate conditional expectations given $\theta$ in general. It is based on Bayes formula~\eqref{eq:BFWR} cast in its expectation form: 
\begin{eqnarray} \label{eq:depart}
 \mathbb{E}_\mu ( \mathcal{O} | \theta ) & = &  \frac{ \int_{\bar{\mathcal{Z}}} \sum_{\ell=0}^L  \mathcal{O}(\tilde{z}_\ell) \mu (\theta, \ell | \bar{z}) \bar{\mathcal{P}}_A ( \mathcal{D} \bar{z}) }{\int_{\bar{\mathcal{Z}}} \sum_{\ell=0}^L \mu (\theta, \ell | \bar{z}) \bar{\mathcal{P}}_A ( \mathcal{D} \bar{z})}. 
\end{eqnarray}
The fact that this expectation is equivalent to the expectation of interest can be checked by plugging 
\begin{equation}
  \sum_{\ell=0}^L \mathcal{O}(\tilde{z}_\ell) \mu (\theta, \ell | \bar{z} ) = \int_\mathcal{Z} \mathcal{O}(z) \mu (\theta, z | \bar{z} ) \mathcal{D}z
\end{equation}
in Eq.~\eqref{eq:depart} and simplifying:  
\begin{eqnarray}
 \mathbb{E}_\mu ( \mathcal{O} | \theta)  & = &  \frac{  \int_{\bar{\mathcal{Z}},\mathcal{Z} } \mathcal{O}(z) \mu (\theta, z | \bar{z}) \bar{\mathcal{P}}_A (\bar{z}) \mathcal{D} \bar{z} \mathcal{D} z  }{ \int_{\bar{\mathcal{Z}},\mathcal{Z}} \mu (\theta, z | \bar{z}) \bar{\mathcal{P}}_A ( \bar{z}) \mathcal{D} \bar{z} \mathcal{D}z }\nonumber \\
     & = &  \frac{  \int_{\mathcal{Z} }   \mathcal{O}(z) \mathrm{p}_A(\theta , \mathcal{D} z ) }{ \int_{\mathcal{Z}} \mathrm{p}_A ( \theta, \mathcal{D} z )  } \\ 
     & = &  \int_{\mathcal{Z} }  \mathcal{O}(z) \pi(\mathcal{D} z | \theta )  = \mathbb{E}_\pi ( \mathcal{O} | \theta ). 
\end{eqnarray}

This procedure is adapted from~\cite{athenes:2012}. For additional references on waste-recycling, see the original articles~\cite{frenkel:2004,frenkel:2006}, articles discussing connections with Bayes formulae~\cite{athenes:2007,athenes:2010}, a mathematical analysis of convergence~\cite{delmas:2009} and some applications~\cite{athenes:2008,adjanor:2011,kim:2011}. 

\end{document}